\documentstyle[epsfig,color,12pt]{article}

\include{epsf}

\newcommand{\EEGG}{\rm e^+ e^-\rightarrow \gamma\gamma(\gamma)}

\def\beq{\begin{equation}}
\def\eeq{\end{equation}}

\begin{document}

\centerline{\large\bf Elementary superconductivity in nonlinear
electrodynamics coupled to gravity}

\vskip 0.1in

\centerline{\large\it Irina~Dymnikova}

\vskip 0.1in

\centerline{\sl A.F. Ioffe Physico-Technical Institute,
Politekhnicheskaja 26, St.Petersburg, 194021 Russia}

\centerline{\sl Department of Mathematics and Computer Science,
University of Warmia and Mazury,}

\centerline{\sl S{\l}oneczna 54, 10-710  Olsztyn, Poland; e-mail:
irina@uwm.edu.pl}

\vskip0.1in

{\bf Abstract}

Source-free equations of nonlinear electrodynamics minimally
coupled to gravity admit regular axially symmetric asymptotically
Kerr-Newman solutions which describe charged rotating black holes
and electromagnetic spinning solitons (lumps). Asymptotic analysis
of solutions shows, for both black holes and solitons, the
existence of de Sitter vacuum interior which has the properties of
a perfect conductor and ideal diamagnetic and displays
superconducting behaviour which can be responsible for practically
unlimited life time of the electron. Superconducting current flows
on the equatorial ring replacing the Kerr ring singularity of the
Kerr-Newman geometry. Interior de Sitter vacuum supplies the
electron with the finite positive electromagnetic mass related the
interior de Sitter vacuum of the electroweak scale and to breaking
of space-time symmetry, which allows to explain the mass-square
differences for neutrino and the appearance of the minimal length
scale in the annihilation reaction $\EEGG$.

{\bf Journal Reference: Journal of Gravity, vol.2015, Article ID
904171 (2015)}

\section{Introduction}

Quantum electrodynamics applies the point-like idealization for
leptons, which well describes  in- and out- states of particles at
the distances sufficiently large as compared with their eventual
sizes. In experiments on high energy scattering, leptons are found
structureless  down to $\sim{10^{-16}}$ cm. There exist however
experiments, in which particles approach each other so close that
their annihilation results in their complete destruction. Study of
the electromagnetic reaction $\EEGG$ with using the data from
VENUS, TOPAZ, ALEPH, DELPHI, L3 and OPAL, reveals with the $5
\sigma$ significance the existence of the minimal length
$l_e=1.57\times 10^{-17}$ cm  at the scale $E=1.253$ TeV
\cite{us2014}. The annihilation reaction can be a source of
information about possible internal structure of leptons, which
requires an extended model for the electron.

\vskip0.1in

In 1962 Dirac proposed to assume the electron to have a finite
size, with no a priori constraints fixing its size and shape
\cite{dirac}. In his model the electron is visualized as a
spherical shell which serves as a source of electromagnetic field
and is supplied with a cohesive force (Poincar\'e stress) of a
non-electromagnetic origin, needed to prevent the electron from
flying apart under the Coulomb repulsion \cite{dirac}.

\vskip0.1in

The Kerr-Newman solution to the source-free Maxwell-Einstein
equations found in 1965 \cite{newman}
  $$
ds^2 = \frac{(2mr-e^2) - \Sigma}{\Sigma} dt^2  +
\frac{\Sigma}{\Delta} dr^2 + \Sigma d\theta^2 -
\frac{2a(2mr-e^2)\sin^2\theta}{\Sigma}dt d\phi
 $$
 \beq
 + \biggl(r^2 + a^2
+ \frac{(2mr-e^2)a^2\sin^2\theta}{\Sigma}\biggr)\sin^2\theta
d\phi^2; ~~ \Delta = r^2 - 2mr + a^2 + e^2 ,
                                                                        \label{kn}
  \eeq
where $\Sigma$-function and the associated electromagnetic
potential read
  \beq
 \Sigma = r^2 + a^2\cos^2\theta; ~~
 A_i = - \frac{e r}{\Sigma}[1; 0, 0, -a\sin^2\theta] ,
                                                         \label{kn1}
 \eeq
inspired the further search since Carter discovered in 1968  that
the parameter $a$ couples with the mass $m$ to give the angular
momentum $J=ma$ and independently couples with the charge $e$ to
give an asymptotic magnetic dipole moment $\mu=ea$, so that the
gyromagnetic ratio $e/m$ is exactly the same as predicted for a
spinning particle by the Dirac equation \cite{carter}.

At the same time Carter discovered the big trouble of the
Kerr-Newman geometry just in the case appropriate for the
electron, $a^2+e^2>m^2$, when there are no Killing horizons, the
manifold is geodesically complete (except for geodesics which
reach the singularity), and any point can be connected to any
other point by both a future and a past directed time-like curve.
Closed time-like curves originate in the region where
$g_{\phi\phi} <0$, can extend over the whole manifold and cannot
be removed by taking a covering space \cite{carter}.

The source models for the Kerr-Newman exterior fields, involving a
screening or covering of causally dangerous region and Poincar\'e
stress of different origins, can be roughly divided into disk-like
\cite{werner,bur74,hamity,lopez1}, shell-like
\cite{delacruz,cohen,boyer1,lopez}, bag-like
\cite{boyer,trumper,tiomno,bur89,behm,bur2014}, and string-like
(\cite{bur2013} and references therein). The problem of matching
the Kerr-Newman exterior to a rotating material source does not
have a unique solution, since one is free
 to choose arbitrarily the boundary between the exterior
and the interior \cite{werner} as well as an interior model.

The Dirac proposal to approach the electron without a priori
constraints on its size and shape, can be applied in the context
of the Coleman lump (physical soliton)  as a non-singular,
non-dissipative solution of finite energy holding itself together
by its own self-interaction \cite{coleman}. An appropriate
instrument to shed some light on the purely electromagnetic
reaction of $e^{+}e^-$ annihilation, is nonlinear electrodynamics
coupled to gravity (NED-GR)\footnote{NED theories appear as
low-energy effective limits in certain models of string/M-theories
\cite{fradkin,tseytlin,witten}.}.

Nonlinear electrodynamics was proposed by Born and Infeld as
founded on two basic points: to consider electromagnetic field and
particles within the frame of one physical entity which is
electromagnetic field; to avoid letting physical quantities become
infinite \cite{born}. In their theory a total energy is finite,
particles are considered as singularities of the field, but it is
also possible to obtain the finite electron radius by introducing
an upper limit on the electric field \cite{born}.

The Born-Infeld program can be realized in nonlinear
electrodynamics minimally coupled to gravity. Source-free NED-GR
equations admit regular causally safe axially symmetric
asymptotically Kerr-Newman solutions which describe regular
rotating charged black holes and  electromagnetic spinning
solitons (lumps) \cite{me2006,interior,portrait}.

For any gauge-invariant Lagrangian ${\cal L}(F)$, stress-energy
tensor of  electromagnetic field
 \beq
\kappa T^{\mu}_{\nu}=-2{\cal L}_F
F_{\nu\alpha}F^{\mu\alpha}+\frac{1}{2}\delta^{\mu}_{\nu}{\cal
L};~~~\kappa=8\pi G ,
                                                                            \label{set}
  \eeq
  where $F_{\mu\nu}=\partial_{\mu}A_{\nu}-\partial_{\nu}A_{\mu}$
  (greek indices run from 0 to 3) and ${\cal L}_F=d{\cal L}/dF$, in
  the spherically symmetric case
has the algebraic structure
 \beq
T^0_0=T^1_1 ~ ~ ~(p_r=-\rho) .
                       \label{myvac}
  \eeq
  since the only essential components of $F_{\mu\nu}$ are a radial
  electric field $F_{01}$ and a radial magnetic field $F_{23}$.
Regular spherically symmetric solutions with stress-energy tensors
specified by (\ref{myvac}) satisfying the weak energy condition
(non-negativity of density as measured by any local observe), have
obligatory de Sitter center with $p=-\rho$
\cite{me92,me2000,me2002,me2003}. In NED-GR regular spherical
solutions the weak energy condition is always satisfied and de
Sitter vacuum provides a proper cut-off on self-interaction
divergent for a point charge \cite{me2004,me2006}. They can be
transformed  into regular axially symmetric solutions by the
G\"urses-G\"ursey algorithm \cite{gurses,burhild}.

Here we outline the generic properties of regular rotating charged
black holes and solitons.

\section{Basic equations}

Nonlinear electrodynamics minimally coupled to gravity is
described by the action
 \beq
S=\frac{1}{16\pi G}\int{d^4 x\sqrt{-g}[R-{\cal L}(F)]}; ~~ ~
F=F_{\mu\nu}F^{\mu\nu} ,
                                                          \label{action}
 \eeq
where $R$ is the scalar curvature. The Lagrangian ${\cal L}(F)$ is
an arbitrary function of $F$ which should have the Maxwell limit,
${\cal L} \rightarrow F,~ {\cal L}_F\rightarrow 1$ in the weak
field regime.

Variation with respect to $A^{\mu}$ and $g_{\mu\nu}$ yields
 the dynamic field equations \beq \nabla_{\mu}({\cal L}_F
F^{\mu\nu})=0;~~~ \nabla_{\mu}{^*}F^{\mu\nu}=0;~~
^{\star}F^{\mu\nu}=\frac{1}{2}\eta^{\mu\nu\alpha\beta}F_{\alpha\beta};
~~ \eta^{0123}=-\frac{1}{\sqrt{-g}}
                                                                       \label{DynEq}
 \eeq
and the Einstein equations $G_{\mu\nu}=-\kappa T_{\mu\nu}$ with
$T_{\mu\nu}$ given by (\ref{set}).

NED-GR equations do not admit regular spherically symmetric
solutions with the Maxwell center \cite{kirill}, but they  admit
regular solutions with the de Sitter center \cite{me2004}. The
question of correct description  of NED-GR regular electrically
charged structures by the Lagrange dynamics  is clarified in
\cite{us2015}. Regular solutions  satisfying (\ref{myvac}) are
described by the metric
 \beq
 ds^2=g(r) dt^2 - \frac{dr^2}{g(r)} - r^2 d\Omega^2; ~~g(r)=1-\frac{2{\cal M}(r)}{r}; ~~
 {\cal
M}(r)=4\pi\int_0^r{~\tilde\rho(x)x^2dx} ,
                                                                                     \label{spherMetric}
  \eeq
 with the electromagnetic density $\tilde\rho(r)=T^t_t(r)$ from
(\ref{set}). This metric has the de Sitter asymptotic as
$r\rightarrow 0$ and the Reissner-Nordstr\"om asymptotic as
$r\rightarrow\infty$ \cite{me2004}.

The regular spherical solutions generated by (\ref{myvac}) belong
to the Kerr-Schild class \cite{ks,behm,ssqv} and can be
transformed by the G\"urses-G\"ursey algorithm \cite{gurses} into
regular axially symmetric solutions which describe regular
rotating electrically charged objects, asymptotically Kerr-Newman
for a distant observer  \cite{burhild,me2006}.

In the Boyer-Lindquist coordinates the rotating metric reads (in
the units $G=c=1$) \cite{gurses}
  \beq ds^2 =
\frac{2f(r) - \Sigma}{\Sigma} dt^2  + \frac{\Sigma}{\Delta} dr^2 +
\Sigma d\theta^2 - \frac{4af(r)\sin^2\theta}{\Sigma}dt d\phi +
\biggl(r^2 + a^2 +
\frac{2f(r)a^2\sin^2\theta}{\Sigma}\biggr)\sin^2\theta d\phi^2 ,
                                                                                \label{metric}
  \eeq
 where $\Delta = r^2 + a^2 - 2f(r)$.
 A function $f(r)=r{\cal M}(r)$ comes from a spherically
symmetric solution \cite{gurses}. For the Kerr-Newman geometry
$2f(r) = 2mr - e^2$ is responsible for causality violation related
to regions where $g_{\phi\phi} < 0$ in (\ref{kn}). For
 NED-GR regular solutions satisfying the weak energy condition,
${\cal M}(r)$ is non-negative function  monotonically growing from
$4\pi\tilde\rho(0)r^3/3$ as $r\rightarrow 0$ to $m-e^2/2r$ as
$r\rightarrow\infty$ \cite{me2004}. This guarantees the causal
safety on the whole manifold due to $f(r)\geq 0$ and $g_{\phi\phi}
> 0$ in (\ref{metric}).

The coordinate $r$ is defined as an affine parameter along either
of two principal null congruences, and
 the surfaces of constant $r$ are the oblate
confocal ellipsoids of revolution
 \beq
r^4-r^2(x^2+y^2+z^2-a^2)-a^2 z^2=0 ,
                                                                            \label{ellipsoid}
\eeq
 which degenerate, for $r = 0$, to the equatorial disk
 \beq
 x^2 + y^2 \leq a^2, ~ ~~ z = 0 ,
                                                                           \label{disk}
 \eeq
centered on the symmetry axis and bounded by the ring
$x^2+y^2=a^2$ ($r=0, \theta=\pi/2$)  \cite{chandra}.

\section{Geometry}

 The
Cartesian coordinates $x, y, z$ are related to the Boyer-Lindquist
coordinates $r, \theta, \phi$ by
 \beq
 x^2 + y^2 =
(r^2 + a^2)\sin^2\theta;~~z=r\cos\theta .
                                                                           \label{coordinates}
 \eeq
 The anisotropic stress-energy tensor responsible for
(\ref{metric})  can be written in the form \cite{gurses}
     \beq
 T_{\mu\nu}=(\rho + p_{\perp})(u_{\mu}u_{\nu} -
l_{\mu}l_{\nu}) + p_{\perp} g_{\mu\nu}
                                                                              \label{GGset}
  \eeq
in the orthonormal tetrad
 \beq
u^{\mu} = \frac{1}{\sqrt{\pm\Delta\Sigma}}[(r^2 + a^2)
\delta^{\mu}_0 + a \delta^{\mu}_3], ~~ l^{\mu} =
\sqrt{\frac{\pm\Delta}{\Sigma}}\delta^{\mu}_1,  ~~ n^{\mu} =
\frac{1}{\sqrt{\Sigma}}\delta^{\mu}_2,  ~~ m^{\mu} =
\frac{-1}{\sqrt{\Sigma}\sin\theta}[a\sin^2\theta \delta^{\mu}_0 +
\delta^{\mu}_3] .
                                                                              \label{tetrad}
 \eeq
 The sign plus refers to the regions outside the event horizon
 and inside the Cauchy horizon where the vector $u^{\mu}$ is time-like,
 and the sign minus refers to
 the regions between the horizons where the vector $l^{\mu}$ is
 time-like. The vectors $m^{\mu}$ and $n^{\mu}$ are space-like in
 all regions.

 The eigenvalues of the stress-energy tensor (\ref{set}) in the co-rotating frame
 where each of ellipsoidal layers rotates with the angular velocity
$\omega(r) = u^{\phi}/u^t = a/(r^2 + a^2)$ \cite{behm}, are
defined by
  \beq
  T_{\mu\nu}u^{\mu}u^{\nu} = \rho(r, \theta);
~~T_{\mu\nu}l^{\mu}l^{\nu} = p_r = - \rho;  ~~
T_{\mu\nu}n^{\mu}n^{\nu} = T_{\mu\nu}m^{\mu}m^{\nu} = p_{\perp}(r,
\theta) .
                                                                           \label{set-eigenvalues}
 \eeq
in the regions outside the event horizon and inside the Cauchy
horizon where density is defined as the eigenvalue of the
time-like eigenvector $u^{\mu}$. They are related to the function
$f(r)$ as \cite{behm}  as $\kappa\Sigma^2\rho = {2(f'r - f); ~
~\kappa\Sigma^2 p_{\perp} = 2(f'r - f) - f^{\prime\prime}\Sigma}$
\cite{behm}. This gives
 \beq
 \kappa\rho(r, \theta)=\frac{r^4}{\Sigma^2}{\tilde\rho}(r);~~~
\kappa(p_{\perp}+\rho)=2\left(\frac{r^4}{\Sigma^2}-\frac{r^2}{\Sigma}\right)
{\tilde\rho}(r)-\frac{r^3}{2\Sigma}{\tilde\rho}^{\prime}(r)=\frac{2r^2}{\Sigma^2}\left(\frac{\Sigma
r}{4}|\tilde\rho^{\prime}|
  -{\tilde\rho}a^2\cos^2\theta\right) ,
                                                                               \label{press-dens}
  \eeq
where $\tilde{\rho}(r)$ is a relevant spherically symmetric
density profile.  The prime denotes the derivative with respect to
$r$.

{\bf Horizons, ergospheres and ergoregions -} Horizons are defined
by zeros of the function $\Delta(r)$ given by
 \beq
\Delta(r)=r^2 + a^2 - 2f(r) = a^2+r^2g(r) .
                                                                             \label{deltag}
  \eeq
$\Delta=a^2$ at zero points of the metric function $g(r_{h})=0$
and changes from $\Delta=a^2$ as $rt=0$ to
$\Delta\rightarrow\infty$ as $r\rightarrow\infty$.

Ergosphere is a surface of a static limit $g_{tt}=0$ given by
 \beq
 g_{tt}(r, \theta)=r^2 + a^2\cos^2\theta - 2f(r)=0
                                                                             \label{ergosphere}
  \eeq
It follows that $z^2=(2r^2f(r)-r^4)/a^2$. Each point of the
ergosphere belongs to some of confocal ellipsoids
(\ref{ellipsoid}) covering the whole space as the coordinate
surfaces $r$=const. At the $z$-axis equations (\ref{deltag}) and
(\ref{ergosphere}) are identical, so that the minor axis of
ergosphere is equal $r_+$.

For black holes ergoregions (the regions where $g_{tt} < 0$) exist
for any density profile.  Black holes have at most two horizons.
Ergoregions exist between the event horizon and ergosphere.
Solitons are objects without horizons, they can have two, one or
no  ergoregions; this depends on the particular form of a density
profile $\tilde\rho(r)$ and on the values of parameters
\cite{interior}.

{\bf De Sitter vacuum interiors -} Rotation transforms the de
Sitter center to the de Sitter equatorial disk (\ref{disk}) which
exists in each regular axially symmetric geometry.
 In the limit $r\rightarrow 0$, on the disk (\ref{disk}),
$r^2/\Sigma\rightarrow 1$ \cite{me2006}. For the spherical
solutions regularity requires $\tilde\rho(r) < \infty$,
$r{\tilde\rho}^{\prime}(r)\rightarrow 0$ and
$2f(r)\rightarrow\kappa\tilde\rho(0){r^4}/3$ as $r\rightarrow 0$
\cite{me2004}, so that the disk $r=0$ is intrinsically flat
\cite{me2006}. Equation (\ref{press-dens}) gives in this limit
 the equation of state on the disk
  \beq
p_{\perp}+\rho=0  ~~~\rightarrow ~ p_{\perp}=p_r=-\rho ,
                                                                         \label{press-on-disk}
  \eeq
which represents   the rotating de Sitter vacuum \cite{me2006}.

Equation (\ref{press-dens}) implies a possibility of generic
violation of the weak energy condition (WEC) which was reported
for several particular models of regular rotating objects
\cite{behm,neves,bambi,stuchlik}. WEC can be violated beyond the
vacuum surface ${\cal E}(r, z)=0$ on which $p_{\perp}+\rho=0$ and
the right-hand side in (\ref{press-dens}) can change its sign
\cite{interior}. It can be expressed through the pressure of a
related spherical solution,
$\tilde{p_{\perp}}=-\tilde\rho-r\tilde\rho^{\prime}/2$
\cite{me2004}, which gives \cite{interior}
  \beq
  \kappa(p_{\perp}+\rho)=\frac{r|\tilde\rho^{\prime}|}{2\Sigma^2}~{\cal
  E}(r,z)=0;~~{\cal
  E}(r,z)=(r^4-z^2P(r));~~P(r)
  =\frac{2a^2}{r|\tilde\rho^{\prime}|}(\tilde\rho
  -\tilde{p_{\perp}}) .
                                                                 \label{e-surface}
 \eeq
The existence of  vacuum surfaces is directly stipulated by
fulfillment of the dominant energy condition ($\tilde\rho \geq
\tilde{p_k}$) for related spherical solutions. Each vacuum ${\cal
E}$-surface contains the de Sitter disk as a bridge and is
entirely confined within the $r_*$-ellipsoid whose minor axis
coincides with $|z|_{max}$ for the ${\cal E}$-surface
\cite{interior}. The squared width of the ${\cal E}$-surface,
$W^2_{\cal E}=(x^2+y^2)_{\cal E} =
(a^2+|z|\sqrt{P(r)})(1-|z|/\sqrt{P(r)})$.
  For regular solutions
$r\tilde\rho^{\prime}\rightarrow 0$,~ $p_{\perp}\rightarrow -\rho$
as $r\rightarrow 0$ \cite{me2004}, and $P(r)\rightarrow
A^2r^{-(n+1)}$ with the integer $n\geq 0$ as $r\rightarrow 0$. The
function $W_{\cal E}(z)$ has the cusp at approaching the disk and
at least two symmetric maxima between $z=\pm r_*$ and $z=0$
\cite{interior}.

In Fig.1 \cite{interior} ${\cal E}$- surface is plotted  for the
electromagnetic soliton  with the regularized Coulomb profile
\cite{me2004}
 \beq
  \tilde\rho=\frac{q^2}{(r^2+r_q^2)^2};~~r_q=\frac{\pi q^2}{8 m} .
                                                                           \label{profile}
 \eeq
Its width in the equatorial plane $W_{\cal E}=a$ and the height
$H_{\cal E} =|z|_{max}=\sqrt{ar_q}$. For the electron $W_{\cal
E}=\lambda_e/2$, where $\lambda_e\simeq {3.9\times 10^{-11}}$ cm,
~$H_{\cal E}\simeq 0.038\lambda_e$, and $\eta=W_{\cal E}/H_{\cal
E}=\sqrt{a/r_q}\simeq 13,2$.

\begin{figure}[htp]
\centering \epsfig{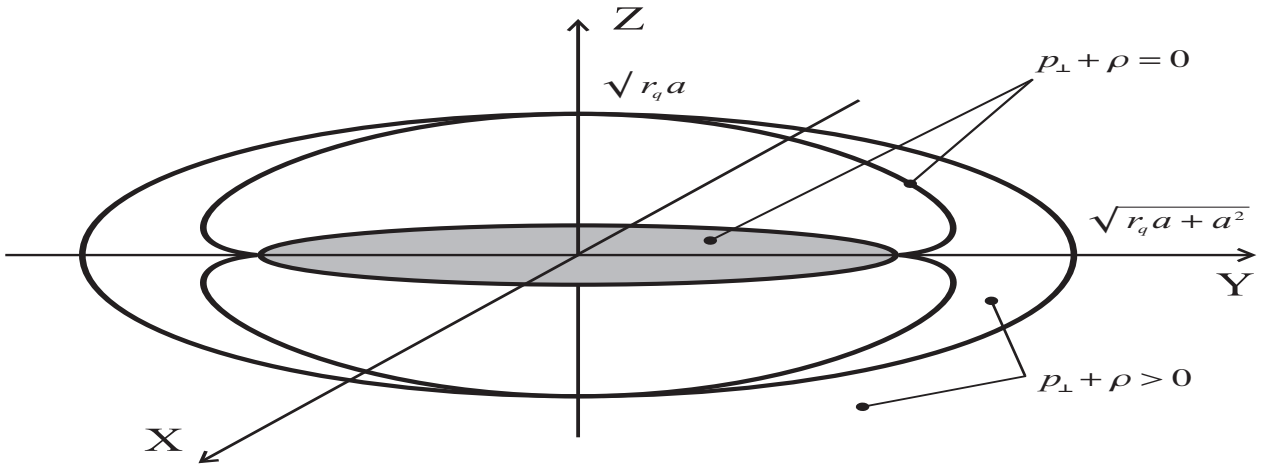}
\caption{Electromagnetic e-lump from nonlinear electrodynamics
coupled to gravity \cite{interior}.}
 \label{fig}
\end{figure}

\section{Electromagnetic fields}

Non-zero field components compatible with the axial symmetry are
$F_{01}, F_{02}, F_{13}, F_{23}$. In geometry with the metric
(\ref{metric}) they are related by
  \beq
F_{31}=a\sin^2\theta F_{10}; ~~ aF_{23}=(r^2+a^2)F_{02} .
                                                                        \label{FieldComp}
  \eeq
The field invariant $F=F_{\mu\nu}F^{\mu\nu}$ in the axially
symmetric case reduces to
  \beq
F=2\left(\frac{F_{20}^2}{a^2\sin^2\theta}-F_{10}^2\right) .
                                                                      \label{AxInvariant}
 \eeq

In terms of the  3-vectors, denoted by latin indices running from
1 to 3 and defined as
   \beq
{E}_j=\{F_{j0}\}; ~~ {D}^j=\{{\cal L}_F F^{0j}\};~
~{B}^j=\{{^*}F^{j0}\};
 ~~ {H}_j=\{{\cal L}_F {^*}F_{0j}\}
                                                                            \label{fields}
  \eeq
the field equations (\ref{DynEq}) take the form of the Maxwell
equations. The electric induction  ${\mathbf{D}}$ and the magnetic
induction $\mathbf{B}$ are related with
 the electric and magnetic field intensities by \cite{me2006}
   \beq
D^{j}=\epsilon^{j}_{k}E^{k}; ~~ B^{j}=\mu^{j}_{k}H^{k} ,
                                                                                 \label{s2}
 \eeq
where $\epsilon_{j}^{k}$ and $\mu_{j}^{k}$ are the tensors of the
electric and magnetic permeability given by \cite{me2006}
  \beq
\epsilon_r^r=\frac{(r^2+a^2)}{\Delta}{\cal L}_F;~~
\epsilon_{\theta}^{\theta}={\cal L}_F; ~~
\mu_r^r=\frac{(r^2+a^2)}{\Delta{\cal L}_F}; ~~
\mu_{\theta}^{\theta}=\frac{1}{{\cal L}_F}  .
                                                                                   \label{s4}
  \eeq
The dynamical equations (\ref{DynEq}) are satisfied by the
functions \cite{me2006}
  \beq
F_{10}=\frac{q}{\Sigma^2{\cal L}_F}(r^2-a^2\cos^2\theta);~
 F_{02}=\frac{q}{\Sigma^2{\cal L}_F}a^2r\sin 2\theta; ~
F_{31}=a\sin^2\theta F_{10}; ~ aF_{23}=(r^2+a^2)F_{02}
                                                                        \label{f6}
 \eeq
in the weak field limit ${\cal L}_F=1$, where they coincide with
the Kerr-Newman fields  \cite{carter,tiomno}, and an integration
constant $q$ is identified as the electric charge. For the
electron $q=-e$, $ma=\hbar/2$ \cite{carter}, $a=\lambda_e/2$,
where $\lambda_e =\hbar/(m_ec)$ is the Compton wavelength. In the
observer region $r\gg\lambda_e$
   \beq
   E_r=-\frac{e}{r^2}\left(1-\frac{\hbar^2}{m_e^2c^2}\frac{3\cos^2\theta}{4r^2}\right);~
   E_{\theta}=\frac{e\hbar^2}{m_e^2c^2}\frac{\sin 2\theta}{4r^3};
   ~~
   B^r=-\frac{e\hbar}{m_ec}\frac{\cos\theta}{r^3}=2\mu_e\frac{\cos\theta}{r^3};
   ~ B_{\theta}=-\mu_e\frac{\sin\theta}{r^4} .
                                                                              \label{lump}
 \eeq
The Planck constant appears here due to discovered by Carter
ability of the Kerr-Newman solution to present the electron as
seen by a distant observer. In terms of the Coleman lump
(\ref{lump}) describes the following situation: The leading term
in $E_r$ gives the Coulomb law as the classical limit $\hbar=0$,
the higher terms represent the quantum corrections.

With taking into account connection of the field components
(\ref{FieldComp}), we have four dynamical equations (\ref{DynEq})
for two field components $F_{01}$, $F_{02}$, and the non-linearity
function ${\cal L}_F$. Condition of compatibility of system of
four equations for three function reduces to the constraint on the
nonlinearity which has the form \cite{interior}
  \beq
\frac{\partial}{\partial r}\bigg(\frac{1}{L_F} \frac{\partial
L_F}{\partial \theta}\bigg)\frac{\partial}{\partial
\theta}\bigg(\frac{1}{L_F} \frac{\partial L_F}{\partial
r}\bigg)+\frac{4a^2\sin^2(\theta)}{\Sigma^2}\frac{1}{L^2_F}\bigg[r\frac{\partial
L_{F}}{\partial r}+\cot(\theta)\frac{\partial L_{F}}{\partial
\theta}\bigg]^2=0 .
                                                                                     \label{cond1}
   \eeq
The functions (\ref{f6}) present asymptotic solutions to the
dynamical equations (\ref{DynEq}) in the limit ${\cal
L}_F\rightarrow\infty$ and ${\cal L}_F\Sigma^2\rightarrow\infty$.
In this limit they satisfy the system of equations (\ref{DynEq})
and the condition of their compatibility (\ref{cond1})
\cite{interior}.

\section{Interior dynamics and elementary superconductivity}

The relation connecting density and pressure with the
electromagnetic fields   reads \cite{me2006}
  \beq
   \kappa (p_{\perp}+\rho)=2{\cal L}_F\left(
F_{10}^2+\frac{F_{20}^2}{a^2\sin^2\theta}\right) .
                                                                                      \label{VIP}
  \eeq
In the limit ${\cal L}_F\rightarrow \infty$  equations (\ref{DynEq}) have asymptotic solutions (\ref{f6})
\cite{me2006,interior}. It results in
\cite{me2006}
    \beq
\kappa{(p_{\perp}+\rho)}=\frac{2q^2}{{\cal L}_F\Sigma^2} .
                                                                         \label{VVIP}
  \eeq
Equation of state on the disk (\ref{press-on-disk}) dictated by
geometry for regular spinning solutions, requires
$p_{\perp}+\rho=0$. It follows ${\cal
L}_F\Sigma^2\rightarrow\infty$ and hence ${\cal
L}_F\rightarrow\infty$, since $\Sigma\rightarrow 0$ on the disk.

The magnetic induction $\mathbf{B}$ vanishes in this limit, so
that $\mathbf{B}=0$ on the disk \cite{me2006}.
 The electric permeability in (\ref{s4})
goes to infinity, the magnetic permeability $\mu={{\cal
L}_F}^{-1}$ vanishes, so that the de Sitter vacuum disk has both
perfect conductor and ideal diamagnetic properties.

In electrodynamics of continued media the transition to a
superconducting state corresponds to the limits
${\mathbf{B}}\rightarrow 0$ and $\mu\rightarrow 0$ in a surface
current
  $
{\mathbf{j_s}}= \frac{(1-\mu)}{4\pi\mu}[{\mathbf{n}}{\mathbf{B}}]$
where ${\mathbf{n}}$ is the normal to the surface;  the right-hand
side then becomes indeterminate, and there is no condition which
would restrict the possible values of the current \cite{landau2}.
Definition of a surface current for a charged surface layer is $4\pi
j_k=[e_{(k)}^{\alpha}F_{\alpha\beta}n^{\beta}]$ \cite{werner},
where $[..]$ denotes a jump across the layer; $e_{(k)}^{\alpha}$
are the tangential base vectors associated with the intrinsic
coordinates on the disk  $t,\phi$, $0\leq\xi\leq\pi/2$;
$n_{\alpha}=(1+q^2/a^2)^{-1/2}\cos\xi~\delta^1_{\alpha}$ is the
unit normal directed upwards \cite{werner}. With using asymptotic
solutions (\ref{f6}) and magnetic permeability $\mu=1/{\cal L}_F$,
we obtain the surface current \cite{portrait}
   \beq
   j_{\phi}=-\frac{q}{2\pi a}
   ~\sqrt{1+q^2/a^2}~\sin^2\xi~\frac{\mu}{\cos^3\xi} .
                                                                                      \label{current}
 \eeq
At approaching the ring $r=0,~\xi=\pi/2$, both terms in the second
fraction go to zero quite independently. As a result the surface
currents on the ring can be any and amount to a non-zero total
value \cite{me2006,portrait}.

The superconducting  current (\ref{current})
replaces the Kerr ring singularity of the Kerr-Newman geometry
and can be considered as a source of the Kerr-Newman fields.
This kind of a source is non-dissipative, so that electrovacuum solitons
present actually e-lumps in accordance with the Coleman definition of the
lump as a non-singular, non-dissipative solution of finite energy
holding itself together by its own self-interaction
\cite{coleman}. Life time of e-lump is unlimited.

De Sitter disk exists in the interior of any regular charged black
hole and soliton.
When a related spherical solution satisfies the dominant energy
condition, it exists as a bridge inside the ${\cal E}$-surface,
defined by $p_{\perp}+\rho=0$. It follows ${\cal
L}_F\rightarrow\infty$ by virtue of  (\ref{VVIP}). The magnetic
permeability vanishes and electric permeability goes to infinity,
so that ${\cal E}$-surface displays the properties of a perfect
conductor and ideal diamagnetic. Also magnetic induction vanishes
on ${\cal E}$-surface by virtue of the asymptotic solutions
(\ref{f6}), so that the Meissner effect occurs there
\cite{me2006,portrait}.  Within ${\cal E}$-surface, in cavities
between its upper and down boundaries and the bridge,  negative
value of $(p_{\perp}+\rho)$ in (\ref{VIP}) would mean  negative
values for electric and magnetic permeabilities inadmissible in
electrodynamics of continued media
 \cite{landau2}. The case, favored by the underlying idea of nonlinearity replacing
a singularity and suggested by vanishing of magnetic induction on
the surrounding ${\cal E}$-surface, is  extension of ${\cal
L}_F\rightarrow\infty$ to its interiors. Then we have de Sitter
vacuum core, $p=-\rho$, with the properties of a perfect conductor
and ideal diamagnetic and magnetic fields vanishing throughout the
whole core, and the weak energy condition is satisfied for regular
rotating charged black holes and solitons.

\section{Summary and discussion}

Nonlinear electrodynamics minimally coupled to gravity admits the
regular axially symmetric solutions asymptotically Kerr-Newman for
a distant observer, which describe regular charged rotating black
holes and electromagnetic solitons.

The basic generic feature of all these objects is the interior de
Sitter vacuum disk with $p=-\rho$ in the co-rotating frame, which
has properties of a perfect conductor and ideal diamagnetic.
Superconducting currents flow on the confining it ring which serve
as a non-dissipative source of the exterior Kerr-Newman replacing
a ring singularity of the Kerr-Newman geometry, and can be
responsible for the practically unlimited life time of the
electron.

In the case when related spherical solution satisfies the dominant
energy condition, de Sitter disk is incorporated as a bridge in
the vacuum ${\cal E}$-surface with the equation of state $p=-\rho$
and properties of the perfect conductor and ideal diamagnetic and
vanishing magnetic induction. This allows to extend these
properties to the interior of the ${\cal E}$-surface, since
otherwise the violation of the weak energy condition in its
interior would lead to the negative values of the electric and
magnetic permeabilities, inadmissible in electrodynamics of
continued media. As a result the weak energy condition is
satisfied for regular rotating objects of this type.

 Mass parameter $m$ in a NED-GR spinning solution, is
the electromagnetic mass \cite{me2004,me2006}, related to interior
de Sitter vacuum and breaking of space-time symmetry from the de
Sitter group for any solution specified by (\ref{myvac})
\cite{me2002}. This has been tested by application of Casimir
invariants of the de Sitter group in the region surrounding the
interaction vertex for the sub-eV particles, which predicts TeV
scale for gravito-electroweak unification and explains the
experimental results known as a negative mass-squared difference
for neutrino \cite{dharam}.

This conforms with the Higgs mechanism for generation of mass via
spontaneous breaking of symmetry of a scalar field vacuum. The
Higgs field is involved in mass generation in its false vacuum
state satisfying $p=-\rho$. Then the space-time symmetry around
the interaction vertex is the de Sitter group, while in the
observer region it is Poincar\'e group (strictly speaking another
de Sitter with much less value of the cosmic vacuum density),
which requires breaking of symmetry in between to the Lorentz
radial boosts only \cite{me2002}. Generation of mass by the Higgs
mechanism must thus involve breaking of space-time symmetry
\cite{portrait}.

Interior de Sitter vacuum  can explain the appearance of
 the minimal length in the reaction $\EEGG$.  The definite feature
 of annihilation process is that at its certain  stage a region of interaction is
neutral and spinless. We can roughly model it by a spherical bag
with de Sitter vacuum interior asymptotically Schwarzschild as
$r\rightarrow\infty$. For such a structure  there exists a zero
gravity surface at which the strong energy condition ($\rho + \sum
p_{k} \geq 0$) is violated and beyond which gravity becomes
repulsive \cite{me96,me2000}. The related length scale
$r_s\simeq{(r_0^2r_g)^{1/3}}$ appears  in any geometry with de
Sitter interior and  Schwarzschild exterior \cite{werner1,me92}.
 Adopting for  de Sitter interior the
vacuum expectation value $v=246$ GeV responsible for the electron
mass \cite{quigg}, we get de Sitter radius $r_0={1.37}$ cm. Radius
$r_s$ at the energy scale $E\simeq{1.253}$ TeV, is
$r_s\simeq{{0.86\times 10^{-16}}}$ cm, so that the scale $l_e =
{1.57\times10^{-17}}$ cm  fits inside a region where gravity is
repulsive. Regular NED-GR solutions provide a  de Sitter cutoff on
electromagnetic self-energy, which can be qualitatively estimated
by $ {e^2}/{r_c^4}\simeq \kappa\rho_0={3}/{r_0^2}$.
 It gives the length scale $r_c$ at which
electromagnetic attraction is balanced by de Sitter gravitational
repulsion $r_c\simeq {1.05\times 10^{-17}}$ cm, sufficiently close
to the experimental value  $l_e$ for such a rough estimate
\cite{us2014}. The minimal length scale $l_e$ can be thus
understood as a distance of the closest approach of annihilating
particles at which electromagnetic attraction is stopped by the
gravitational repulsion due to interior de Sitter vacuum.

\end{document}